\documentclass[12pt]{article}
\usepackage{naturetex}
\usepackage{amsmath}
\begin{document}

\maketitle

\begin{abstract}

Interferometric scattering microscopy has been a very promising technology for highly sensitive label-free imaging of a broad spectrum of biological nanoparticles from proteins to viruses in a high-throughput manner. 
Although it can reveal the specimen’s size and shape information, the chemical composition is inaccessible in interferometric measurements. Infrared spectroscopic imaging provides chemical specificity based on inherent chemical bond vibrations of specimens but lacks the ability to image and resolve individual nanoparticles due to long infrared wavelengths. 
Here, we describe a bond-selective interferometric scattering microscope where the mid-infrared induced photothermal signal is detected by a visible beam in a wide-field common-path interferometry configuration.
A thin film layered substrate is utilized to reduce the reflected light and provide a reference field for the interferometric detection of the weakly scattered field.
A pulsed mid-IR laser is employed to modulate the interferometric signal. 
Subsequent demodulation via a virtual lock-in camera offers simultaneous chemical information about tens of micro- or nano-particles. 
The chemical contrast arises from a minute change in the particle's scattered field in consequence of the vibrational absorption at the target molecule.
We characterize the system with sub-wavelength polymer beads and highlight biological applications by chemically imaging several microorganisms including \textit{Staphylococcus aureus}, \textit{Escherichia coli}, and \textit{Candida albicans}. 
A theoretical framework is established to extend bond-selective interferometric scattering microscopy to a broad range of biological micro- and nano-particles.

\end{abstract}

\section{Introduction}
A light microscope - a common term referring to an optical microscope utilizing visible light to detect and visualize very small objects – is an indispensable tool in a biology laboratory to study biological micro- and nano-particles. 
The size scale of biological particles is a continuum, ranging from 20~nm for the smallest viruses to a few microns for bacteria and cells. 
The biological particles have enormous utility as well as potential adverse impacts in biotechnology, human health, medicine, and food safety~\cite{exosomedrugdelivery,foodborne}.
For example, viruses are the most abundant biological nanoparticles and they have caused deadly outbreaks throughout human history, with the stark and enduring example embodied in the recent global COVID-19 pandemic. 
Their physical and chemical properties strongly influence their biodistribution and interactions with cells~\cite{virusShape}.
Optical microscopy of individual biological particles has significantly improved our understanding of biological, chemical, and physical processes in life sciences~\cite{BNPApplications}. 
While light microscopy is sufficient to observe biological microparticles such as bacteria, its ability to study individual sub-wavelength nanoparticles is limited by their weakly-scattering characteristics due to exceptionally small size and low-refractive-index difference with the surrounding medium.
Elastically scattered light intensity induced by the illumination of the nanoparticles scales with the sixth power of the particle size resulting in very low optical contrast which makes them indistinguishable from the background in scattered intensity-based optical imaging and detection techniques. 
As an indirect method, fluorescence microscopy has been traditionally used to overcome this weak contrast by detecting exogenous or endogenous fluorophores. 
It has led to several breakthroughs in the super-resolved imaging of nanoscale specimens~\cite{superresolution}. 
Yet, fluorescence labels suffer from phototoxicity, photobleaching, photostability, and saturation. 
These severely limit observation of the molecular/nanoparticle dynamics at very short time scales. 
Moreover, fluorescence labeling may perturb sample functionality and structure further emphasizing the need for biological imaging and characterization without any labels and modifications in their natural environment. 
Therefore, it is highly desirable to advance light microscopy to provide sufficient resolution to study the entire size range of biological processes at the micro- and nano-scale~\cite{labelfreemotivation}. 
Furthermore, augmenting light microscopy with compatible characterization techniques to achieve chemical specificity would provide an invaluable capability especially if both microparticles (bacteria) and nanoparticles (viruses) can be studied. 
Here, we exploit interferometric microscopy to develop such a capability. 
While the present experimental work focuses on bacteria characterization, we establish a theoretical framework and requirements to study single biological nanoparticles such as viruses. 

Over the last two decades, wide-field interferometric microscopy techniques have successfully demonstrated high-throughput, sensitive, and fast direct imaging of biological particles down to a single protein level~\cite{ ISCATdifferential, ISCATMass, ISCATproteinSandoghdar, ISCATvirus, DaaboulNL,DaaboulAN,COBRIViral, DaaboulSR, AygunSR, ISCATMicrotubules, neginBacteria}.
These interferometric techniques rely on the coherent detection of the elastically scattered field – rather than its square or intensity – mixing with a strong reference field in a common-path interferometry configuration. 
Thus, interferometric detection enjoys a reduced size dependence of optical contrast (scaling with the third power of the particle size) and enables a significantly enhanced nanoparticle visibility. 
Interferometric scattering (iSCAT)~\cite{ISCATdifferential} microscopy detects nanoparticles captured on a glass coverslip in reflection mode.
iSCAT illuminates the sample with a highly coherent light source, that is laser, to increase the interferometric contrast~\cite{ISCATNatureProtocol}.
Similarly, coherent brightfield microscopy (COBRI) uses the laser illumination in transmission mode followed by background attenuation in the pupil plane to reveal high interferometric contrast~\cite{COBRIpupil}. 
Unlike the laser-based interferometric microscopy methods, single-particle interferometric reflectance imaging sensor (SP-IRIS)~\cite{OguzhanReview} uses a layered silicon substrate with a partially coherent light source, that is a light-emitting diode (LED). 
These common-path interferometry based wide-field microscopy techniques have demonstrated several applications for imaging, detection, and counting of proteins~\cite{ISCATdifferential, ISCATMass,ISCATproteinSandoghdar}, viruses~\cite{ISCATvirus,DaaboulNL,DaaboulAN,COBRIViral}, exosomes~\cite{DaaboulSR,AygunSR}, microtubules~\cite{ISCATMicrotubules}, and metallic nanoparticles~\cite{sevenlerAN, ISCAT3Dtracking, COBRIpupil}.
However, molecular specificity in these interferometric studies is limited to the surface affinity of assays.

Vibrational spectroscopic imaging methods have enabled molecular fingerprinting of molecules with high chemical specificity~\cite{ChengScienceReview}. 
Infrared absorption and inelastic Raman scattering have been utilized in various optical sensing methods to probe spectroscopic signatures of chemical bonds. 
Raman scattering microscopy techniques rely on a non-linear scattering process to achieve diffraction-limited-resolution~\cite{RamanScattering, SRS}. 
Giving complementary information with Raman scattering, infrared (IR) absorption is another type of vibrational spectroscopy method that is routinely used in analytical fields. 
However, IR-based methods have limited applications in resolving nanoparticles since the spatial resolution is typically several micrometers due to long excitation wavelengths in the mid-IR region (2.5-25~µm)~\cite{FTIR1, FTIR2}. 
As a tip-based approach, atomic force microscope infrared spectroscopy (AFM-IR) techniques can resolve structures at the nanoscale (20 nm) but its applications are typically limited to dry samples due to contact requirement~\cite{AFMIR1, AFMIR3}. 
Recently developed mid-infrared photothermal (MIP) microscopy has demonstrated optical detection of photothermal effect induced by IR absorption of the specimen using a visible probe beam~\cite{Delong2016}. 
In principle, the MIP contrast mechanism is similar to those reported in the photothermal microscopy field\cite{OrritPT,OrritRoomtemp,lounis, cichos}.
This emerging technique has been recognized widely in the chemical imaging field and continuously evolved in various geometries including scanning-based~\cite{Delong2016, Zhongming2017, Masaru1ContrastMechanism, Masaru3Review, PanagisOE} and wide-field~\cite{Yeran2019,phase2019,japanPhaseContrastSR,japanQPIOL,japanQPIOptica,Rohit_PNAS_2020} illumination. 
A wide range of applications spanning from material science to life science has been reported by scanning MIP techniques~\cite{miragePharma,mirageNeuron, Masaru2Material,Masaru4Solar,epi2017, IRaman,YurdakulAC,PanagisBOE}.

Although providing great sensitivity to characterize specimens down to single viruses~\cite{YurdakulAC}, the scanning MIP approaches have inherent limitations associated with scanning itself such as inefficient use of IR photons due to IR/visible foci mismatch and low throughput due to the limited imaging speed.
Bai \textit{et al.}~\cite{Yeran2019} has overcome these challenges and demonstrated a wide-field MIP microscopy approach by probing interferometric reflectance change from the sample. 
To achieve camera-based photothermal detection using very fast complementary metal-oxide-semiconductor (CMOS) cameras, a virtual lock-in camera technique has been introduced. 
Camera-based MIP systems have been already applied to extend various label-free imaging methods including low-coherence interference microscopy~\cite{Rohit_PNAS_2020}, quantitative phase imaging~\cite{phase2019,japanPhaseContrastSR,japanQPIOL,japanQPIOptica,ADRIFT}, and dark-field imaging~\cite{yurdakulDF} and led to applications in histopathology and living cells. 
However, individual nanoparticle detection in wide-field MIP configuration remains still challenging due to the very weak scattered light from nanoparticles. 
As a consequence, single nanoparticles become invisible under the strong background from the illumination. 

In this study, we show bond-selective interferometric scattering microscopy for biological nanoparticle fingerprinting. 
We utilize a layered silicon substrate to increase interferometric contrast by reducing background and enhancing the scattered field~\cite{oguzhanOptica, SevenlerBOE}. 
By probing the mid-IR absorption-induced photothermal signal, this study enables chemical specificity beyond the surface affinity. 
Our technique provides vibrational spectroscopic sample information which is not accessible in the previous wide-field interferometric scattering microscopy approaches. 
A theoretical framework for the interferometric photothermal contrast mechanism is discussed. 
The system performance is evaluated using poly (methyl methacrylate) (PMMA) film and beads. 
To show chemical imaging of biological nanoparticles, we provide examples of \textit{Staphylococcus aureus} (\textit{S. aureus}), \textit{Escherichia coli} (\textit{E. coli}), and \textit{Candida albicans} (\textit{C. albicans}). 
Collectively, these results show promise for high-throughput and sensitive label-free imaging of a broad size range of individual bio-nanoparticles, including viruses and exosomes, with high chemical specificity. 

\section{Materials and methods}
\begin{figure*}[!ht]
	\begin{center}
	\includegraphics[width =.9 \textwidth]{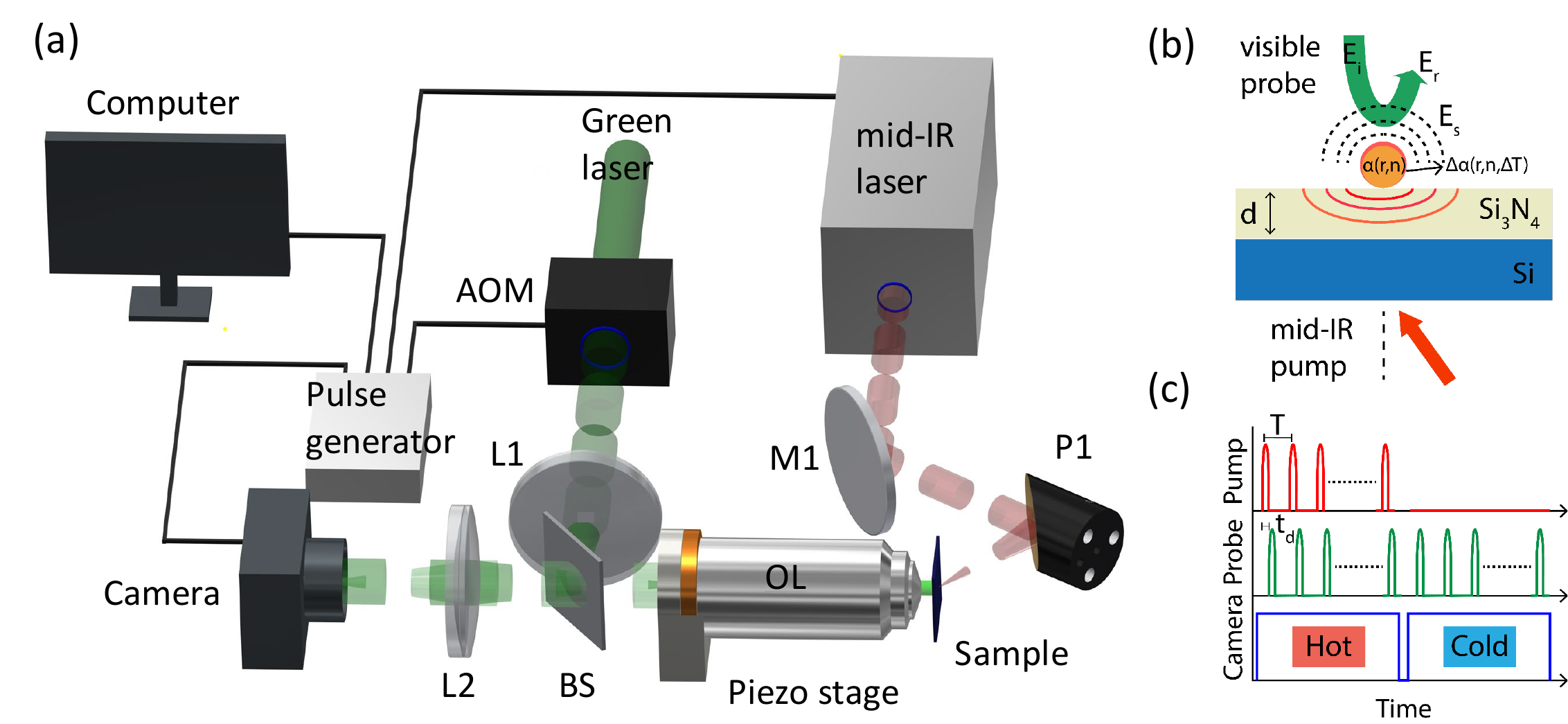}
	\caption{\textbf{Concept and schematic of a bond-selective interferometric scattering microscope.} (a) Schematic setup of wide-field interferometric mid-infrared photothermal microscopy. L1-L2: achromatic doublets; BS: beam-splitter; M1: gold mirror; P1: parabolic gold mirror; AOM: acousto-optic modulator; OL: objective lens. (b) Layout of the common-path interferometric detection, the nanoparticle is placed on top of a 70~nm Si\textsubscript{3}N\textsubscript{4}/Si layered substrate. The visible incidence field ($\mathbf{E_i}$) scatters from the sample ($\mathbf{E_s}$) and reflects off the substrate surface ($\mathbf{E_r}$). The mid-IR pump beam absorbed by the nanoparticle causes temperature rise ($\Delta T$) and induces a change of the particle's polarizability ($\Delta \alpha$) which is a function of the particle's size and refractive index. The pump is incident on the sample with an oblique angle ($\theta = \sim 62 ^\circ$) close to Brewster's angle to improve the IR transmission and avoid absorption by the OL. (c) Virtual lock-in camera detection. The synchronization and timing are controlled by a pulse generator which externally triggers the pump, probe, and camera. $T$ period of the IR and pump pulses is set to 200 kHz. The camera frame rate is set to 200 Hz. The pump beam is modulated at 100 Hz such that odd frames are IR-on (Hot) and even frames are IR-off (Cold). Time delay $t_d$ between the pump and probe pulses is controlled to measure the transient thermal response from particles and maximize the photothermal signal.}
    \label{fig:fig1}
	\end{center}
\end{figure*}

\subsection{Experimental setup}
The schematic of our bond-selective interferometric scattering microscopy is illustrated in figure~\ref{fig:fig1}(a). 
The green pump beam is frequency-doubled from a femtosecond (fs) laser (Chameleon, Coherent) operating at 80~MHz with 1040~nm wavelength and 100~fs pulse width by the second-harmonic generation (SHG) process using a non-linear crystal. 
The second-harmonic generated laser pulses have a 520~nm central wavelength with a 9~nm bandwidth, reducing speckles in the images owing to the low temporal coherence length. 
Before the SHG process, the fs beam is chopped by an acousto-optical modulator (AOM, R23110-0.5, Gooch and Housego) at 200 kHz with a 50\% duty-cycle square waveform. 
To provide wide-field sample illumination, the green laser is focused on the microscope objective's (CFI TU Plan Fluor EPI 50$\times$, NA~0.8, Nikon) back-pupil plane in K\"{o}hler illumination configuration using a field lens (L1) with a 75~mm focal length. 
This illuminates the sample of interest with nearly a plane wave at an incident angle defined by its location at the objective back-pupil. 
The objective lens is mounted on an objective piezo scanner (MIPOS 100 SG RMS, Piezosystem Jena) to adjust the objective nominal focus and acquire defocus image scans. 
To keep the substrate stable in the vertical orientation, the substrate is held by a custom-built vacuum chuck mounted on a tip/tilt kinematic mount. 
The sample is translated by a manual XYZ stage with differential adjusters (PT3A/M, Thorlabs). 
To achieve normal incidence on the sample, we ensure that the green laser is focused at the center of the back-pupil by adjusting two tip/tilt mirrors before the field lens. 
The normal incidence realizes a radially symmetric point spread function (PSF)~\cite{yurdakulAN}. 
We utilize this fact to fine-tune the mirrors by imaging spherical beads. 
The epi-illuminated sample is then imaged onto a 5.0~MP monochromatic area-scan CMOS camera (GS3-U3-51S5M-C, FLIR) through a tube lens (L2). 

A pulsed mid-IR tunable quantum-cascaded laser (QCL, MIRcat, Daylight solutions) illuminates the substrate from the backside at an oblique angle of $\sim$61\textsuperscript{$\circ$}. 
The collimated IR beam is focused on the substrate's front surface using a parabolic gold mirror (P1). 
There are two main reasons for using the oblique IR illumination: (1) it avoids the IR absorption by the objective lens which creates random signal fluctuations and (2) it can improve the IR transmission that enhances the photothermal signal. 
To increase the IR transmission in the silicon substrate, we utilize Brewster's angle in which \textit{p}-polarized light reflection reduces to nearly zero. 
To do so, we topologically rotate the polarization state of the factory default IR beam from \textit{s}-polarized to \textit{p}-polarized which has been detailed in \cite{yurdakulDF}. 
Briefly, we use two 45\textsuperscript{$\circ$} gold mirrors in a periscope configuration in which two mirrors point to each other at a 90\textsuperscript{$\circ$} axial rotation such that the incident beam along the x-axis propagates at the y-axis. 
The first mirror reflects the light upwards and upon this reflection the polarization state changes from \textit{s} to \textit{p}. 
The second mirror preserves its polarization state relative to the optical axis of the substrate. 
The backside oblique IR illumination approach overall improves the IR transmission by about four-fold compared to that of the initial \textit{s}-polarized beam, providing higher IR power on the sample.

\subsection{Interferometric photothermal contrast mechanism}
\label{subsec:theory}
Our bond-selective wide-field interferometric microscopy utilizes common-path interferometry configuration for highly sensitive and stable coherent detection of scattered light from nanoparticles. 
It uses a 70~nm thin layered Si\textsubscript{3}N\textsubscript{4}/Si substrate as depicted in figure~\ref{fig:fig1}(b). 
The nitride film thickness $d$ is nearly $\sim \lambda_{Si_3N_4}/4$ where $\lambda_{Si_3N_4}$ is the wavelength of the light propagating in the nitride layer. 
This specific layer thickness not only reduces the reflected light through destructive interference but also enhances the total scattered light through the constructive self-interference of the forward and backward scattered light~\cite{oguzhanOptica}. 
In this configuration, the incident field $\mathbf{E_i}$ scatters from the sample $\mathbf{E_s}$ and reflects off from the layered substrate $\mathbf{E_r}$ with a complex reflection coefficient dictated by thin-film effect. 
The resulting total driving field $\mathbf{E_d}$ becomes the coherent sum of the complex incident and reflected fields, $\mathbf{E_d  = E_i + E_r}$. The scattered field amplitude $|E_s|$ scales with the total driving field $\mathbf{E_d}$ and particle polarizability $\alpha$ in the dipole limit where the particle size is much smaller than the illumination wavelength~\cite{novotnyBook}. In the dipole approximation ($r\mathrm{\ll \lambda}$), the polarizability of a spherical nanoparticle is given as follows~\cite{polarizabilityBook}:

\begin{equation}
\alpha = 4\pi \mathrm{r}^3\epsilon_m \frac{\epsilon_p - \epsilon_m}{\epsilon_p + 2\epsilon_m},
    \label{eq:polarizability}
\end{equation}

\noindent where r is the particle's radius, $\epsilon_p$ is the particle's dielectric constant, and $\epsilon_m$ is the medium's dielectric constant. Equation~\ref{eq:polarizability} indicates that scattered light intensity is a function of the particle's volume square ($V^2 \propto$ r\textsuperscript{6}) and the refractive index difference between the particle and its surrounding medium. 
The volume dependence of the scattered signal has strong implications on the pure scattering intensity-based detection (e.g., dark-field~\cite{darkfield}) of very small nanoparticles since the signal rapidly falls off as a result of r\textsuperscript{6} dependence. 
Nevertheless, the interferometric detection reduces this square dependence to a linear detection of the scattered field such that the measured signal scales with r\textsuperscript{3} instead. 
In the interferometric microscopy, the detector captures the intensity signal from the coherent sum of reference and scattered fields $I_{det} = |\mathbf{E_r} +\mathbf{E_s}|^2$ as follow:

\begin{equation}
    I_{det} = |E_{r}|^2 + |E_{s}|^2 + 2|E_{r}||E_{s}|\cos(\theta).
    \label{eq:Idet}
\end{equation}

The first term denotes the reflected field intensity $I_{r} = |E_{r}|^2$, the second term denotes the scattered field intensity $I_{s} = |E_{s}|^2$, and the last term denotes the interference signal $2Re \{\mathbf{E_r}\mathbf{E_s}^*\}$. 
The phase term $\cos(\theta)$ denotes the phase difference between complex reflected and scattered fields. 
This phase term has strong implications on signal constituents as it allows for accurate particle size information~\cite{SevenlerBOE}, dielectric characteristics~\cite{oguzhanOE, OguzhanAO}, and its axial position~\cite{ISCAT3Dtracking}. 
For the sake of brevity, we take the phase term as 1 in our formulation, providing the maximized interference signal~\cite{COBRIpupil}. 
Due to the r\textsuperscript{6} dependence of the scattering intensity, the nanoparticles of interest scatter very weakly. 
Under this approximation, the scattered intensity term becomes negligible compared to the reference intensity. 
Therefore, the detected signal can be approximated as $I_{det} \approx |E_{r}|^2 + 2|E_{r}||E_{s}| $. 
This enables the coherent detection of the weak scattering signal with a strong reference. 
The reflected field intensity constitutes the direct component intensity contribution which can be minimized by simple background subtraction. 
Here, we define interferometric signal ($S$) as the background ($B$) subtracted intensity signal as follows:

\begin{equation}
	S  = I_{det} - I_r =  2|E_{s}||E_{r}|.
    \label{eq:intSignal}
\end{equation}

The interferometric detection realizes the linear detection of the scattered field which is proportional to r\textsuperscript{3} whereas the scattering intensity is proportional to r\textsuperscript{6}. 
This leads to high signal-to-noise-ratio (SNR) imaging of small scatterers that generate subtle light scattering contrast above the strong background signal.
We further define the interferometric image contrast as the ratio of the interferometric signal to the background intensity~\cite{iSCATReviewNL},

\begin{equation}
    S_c  = \frac{S}{B} = \frac{I_{det} - I_r}{I_r} =  2\frac{|E_{s}|}{|E_{r}|}.
    \label{eq:intContrast}
\end{equation}

The photothermal signal in our system is measured by probing the change of this interferometric signal. 
The IR absorption increases the temperature at the nanoparticle's vicinity. This temperature rise $\Delta T$ induces a change in the particle's size and refractive index depending on the particle's linear thermal-expansion $\mathrm{\beta_r}$ = (1/r)(dr/dT) and thermo-optic $\mathrm{\beta_n}$ = (1/n)(dn/dT) coefficients. 
As a result, the nanoparticle's polarizability and scattered field amplitude change whereas the reference field is untouched. 
This process is referred to the photothermal effect that causes the temperature-dependent signal change.
Our formulation, in principle, is similar to the previously reported photothermal signal theory based on the scattered field from medium fluctuations~\cite{lounis,cichos}. 
Using equation~\ref{eq:intSignal}, we can express the signal difference $\Delta S$ between IR-on ($S^{hot}$) and IR-off ($S^{cold}$) states as follows: 

\begin{equation}
    \begin{aligned}
    \Delta S &= S^{hot} - S^{cold} = 2|E_r|\Delta |E_s| \\
    \textrm{and}\\ 
    \Delta |E_s| &= |E_s^{hot}| - |E_s^{cold}|,
    \end{aligned}
    \label{eq:PTsignal}
\end{equation}

\noindent where $E_s^{hot} = E_s(T_0+\Delta T)$ and $E_s^{cold} = E_s(T_0)$ are respectively the scattered fields at the IR-on and IR-off states with the pre-IR pulse temperature of $T_0$.
Equation~\ref{eq:PTsignal} implies that interferometric photothermal signal is detected through the linear detection of the scattering amplitude change with a strong reference field~\cite{COBRIPhotothermal}. 
This has a similar detection principle to that of the interferometric signal (see equation~\ref{eq:Idet}). 
Analogous to equation~\ref{eq:intContrast}, the photothermal image contrast can be expressed as the interferometric contrast difference at two states. 

\begin{equation}
    \Delta S_c = S_c^{hot} - S_c^{cold} =  2\frac{\Delta|E_{s}|}{|E_{r}|} = S_c\frac{\Delta|E_{s}|}{|E_{s}^{cold}|}.
    \label{eq:PTcontrast}
\end{equation}

We measure the photothermal signal as the intensity difference at the camera. 
To generalize the photothermal signal quantification, we define the intensity modulation fraction which can be expressed as the ratio of the photothermal signal to the interferometric signal at the pre-pulse state $M_{PT} = \Delta S/ S^{cold}$. 
By plugging equation~\ref{eq:polarizability}, the modulation fraction can be calculated as follows~\cite{YurdakulAC}:

\begin{equation}
    M_{PT} = \frac{\Delta S} {S^{cold}} = \frac{\Delta S_c} {S_c^{cold}} = \frac{\Delta|E_{s}|}{|E_{s}^{cold}|} \propto \frac{\Delta \alpha}{\alpha} \\ \approx 3\Delta T \left( \beta_r + \frac{2\epsilon_p\epsilon_m}{(\epsilon_p + 2\epsilon_m)(\epsilon_p - \epsilon_m)}\beta_n \right ).
    \label{eq:PTmod}
\end{equation}

The equations above indicate that the photothermal signal contrast scales with the modulation fraction and interferometric contrast. 
Therefore, a reference signal reduction through either substrate engineering~\cite{oguzhanOE} or pupil engineering~\cite{oguzhanOptica, kukuraPupil, COBRIpupil} could enhance the photothermal signal visibility. 
The photothermal modulation is typically in negative five/six orders of magnitude for a given 1K temperature rise. 
For example, PMMA has a linear thermal expansion coefficient of $\mathrm{\beta_r}$ = 90 $\times$ 10\textsuperscript{-6} K\textsuperscript{-1}~\cite{thermalExpansion} and thermo-optic coefficient of $\mathrm{n_{PMMA}\beta_n}$ = -1.1 $\times$ 10\textsuperscript{-4} K\textsuperscript{-1}~\cite{PMMAthermooptic} with a refractive index of n\textsubscript{PMMA} = 1.49~\cite{PMMArefractiveindex} at 520~nm wavelength. 
The modulation fraction for a PMMA bead surrounded in the air then becomes nearly 0.01\% at $\Delta T$ = 2K. 

\subsection{Camera-based photothermal detection mechanism}

To measure the photothermal signal in the wide-field system, we employ the virtual lock-in camera detection reported in~\cite{Yeran2019}. 
As shown in figure~\ref{fig:fig1}(c), while a pulse train of visible beam continuously illuminates the sample, the IR beam simultaneously illuminates the sample at every other camera frame. In other words, the IR beam is turned on and off at the subsequent frames. 
The IR pulse period of 5~µs provides enough time to reach steady-state cooling for nanoparticles which have typically fast cooling time smaller than 1~µs\cite{yurdakulDF}.
We refer the IR-on and IR-off frames to \textit{``hot''} and \textit{``cold''}, respectively. 
The photothermal image is then obtained by taking the difference image, subtracting the cold frames from the hot frames. 
\textit{t\textsubscript{d}} denotes the time delay between individual IR and visible pulses. 
The camera, visible, and IR beams are synchronized by a pulse generator master clock which externally triggers each instrument. 
The pulse generator (9254-TZ50-US, Quantum composers) outputs three pulse waves with different duty cycles, pulse widths, and time delays.
This allows for individual control over the AOM, camera exposure, and mid-IR QCL. 
The AOM is triggered at 200~kHz with a 200~ns pulse width to obtain the 200~ns visible pulse train. 
The camera is triggered at 200~Hz using the pulse generator output in a duty cycle mode at 1 on and 999 off. 
To obtain the hot and cold frames, the mid-IR QCL laser is externally triggered at 200 kHz using the duty cycle mode at 1000 on and 1000 off. 
The IR-pulse width is set to 1~µs using the QCL's internal settings. 
The time delay \textit{t\textsubscript{d}} was set to $\sim$500~ns to maximize the photothermal signal. \textit{t\textsubscript{d}} is empirically determined by measuring the transient photothermal response via time-gated pump-probe approach~\cite{phase2019}. 
For the particle of interest, the photothermal signal peaks are around the same time delay of $\sim$500~ns. 
Therefore, the time delay remains constant during the experiments unless otherwise noted.

The image acquisition and the piezo scanner and pulse generator controls are implemented on a custom-built Python script.
For acquisition, we use an open-source Python module named ``PySpin'' provided by the camera company. 
The camera captures the frames sequentially at each trigger pulse and transfers the captured images to the computer via the universal serial port.
Each image is self-normalized by the average intensity at a predetermined region of interest outside of the IR spot. 
This minimizes the effect of possible visible laser intensity fluctuations across multiple frames. 
Even/odd-numbered frames are then temporally stored in the memory and directly averaged into a single 2D array at their corresponding data arrays as \textit{``hot''} and \textit{``cold''}, respectively.
A large number of frame averaging is required to obtain high SNR that can distinguish the photothermal signal from the background noise.
We typically acquire 10000 frames in total, that is, 5000 hot and 5000 cold and average them in real-time. 
This computationally efficient image acquisition approach significantly reduces the memory and space requirement for a large number of frames ($O (N) \rightarrow O (1)$). 
In other words, only two averaged hot and cold images (a few MBs) are saved on the disk at the end of each photothermal image acquisition, instead of saving gigs of image data to average at the post-process. 
We note that all images can also be saved for SNR characterization experiments.

\subsection{Sample preparation}
A 100~mm double side polished 100~nm low-pressure chemical vapor deposition (LPCVD) Nitride (Si\textsubscript{3}N\textsubscript{4}) on silicon (Si) wafer with 500~µm thickness (University Wafer) was first etched down to $\sim$70~nm.
A photolithography process was then used to pattern reference regions for easy focus find, followed by dicing to 10~mm $\times$ 20~mm rectangular chips. Half of the substrate interfaces with the vacuum chuck, the remaining part is used for the sample immobilization. 
A stock solution of 500~nm PMMA nanospheres (MMA500, Phosphorex) was diluted with deionized water by 1:10, followed by a spin coating on the nitride substrate. 
In bacteria experiments, we used two strains of bacteria \textit{S. aureus} ATCC 6538 and \textit{E. coli} BW 25113. 
These strains were obtained from the American Type Culture Collection (ATCC) and the Biodefense and Emerging Infections Research Resources Repository (BEI Resources), respectively. 
To prepare bacteria samples for chemical imaging, the bacterial strains were first cultured in cation-adjusted Mueller-Hinton Broth (Thermo Fisher Scientific) media and grown to the logarithmic phase. 
1~mL of bacteria sample was centrifuged, washed twice with purified water, and then fixed by a 10\% (w/v) formalin solution (Thermo Fisher Scientific), and then were washed twice with purified water. 
To load the bacteria sample, 2~µL of either \textit{S. aureus} or \textit{E. coli} bacteria solution was incubated on the substrate at room temperature until the surface dried. 
In fungi experiments, the live \textit{C. albicans} were dispensed in PBS and dropped on the silicon substrate. To avoid motion artifacts from water fluctuations, another thin piece of the coverslip was used on top.  

\section{Results and Discussion}
\begin{figure}[!ht]
	\begin{center}
	\includegraphics[width =.6 \textwidth]{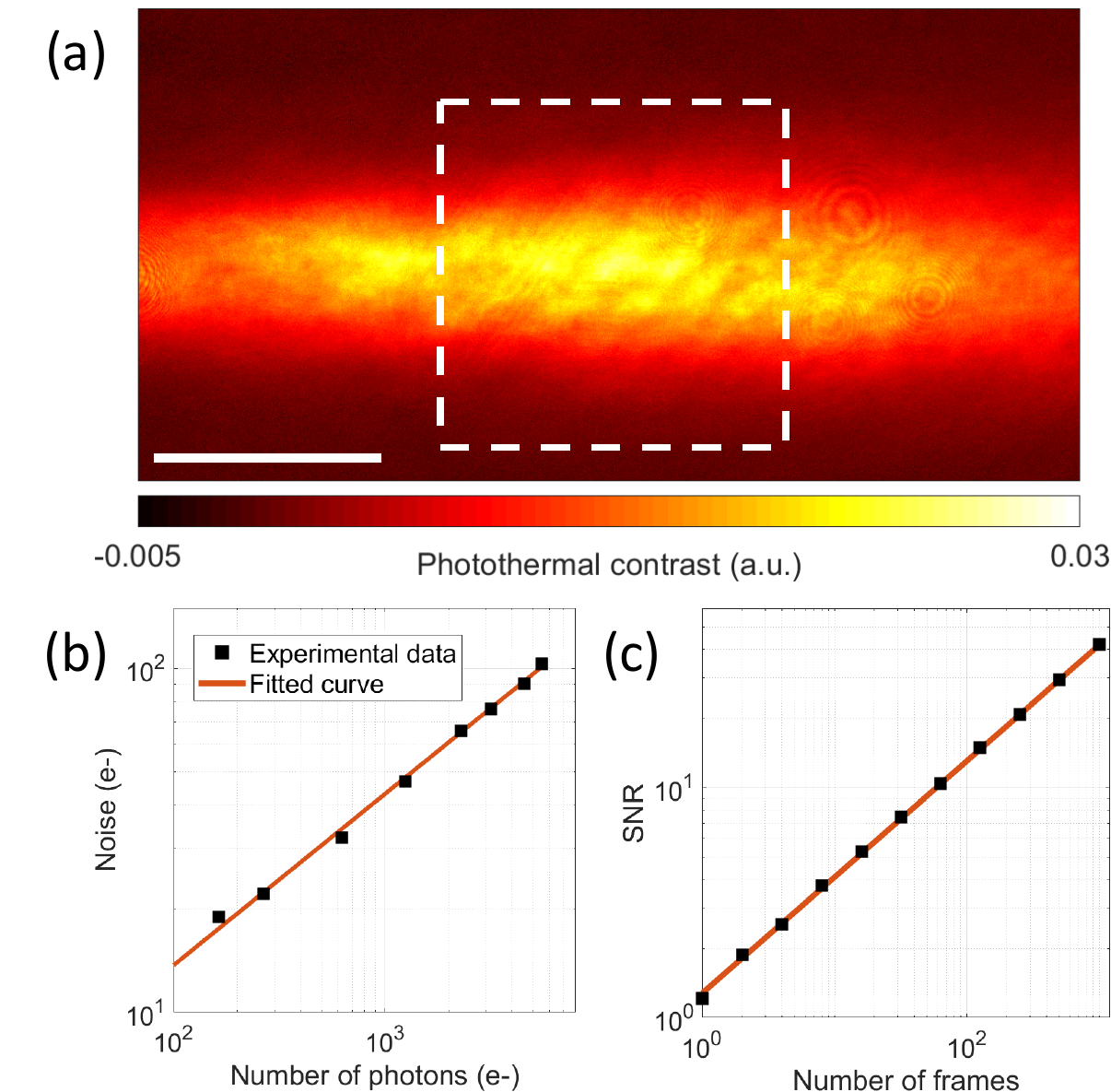}
	\caption{\textbf{Bond-selective interferometric imaging of PMMA thin-film on top of a silicon substrate and system SNR characterization} (a) Interferometric MIP images at C=O stretching absorption peak (1729~cm\textsuperscript{-1}). The dashed square indicates the region of interest in the analyzed photothermal images. (b) Calculated noise in a single photothermal image for different collected photons (\textit{P}). The fitted curve ($\alpha P^{0.498}$) has a slope of 0.498 in logarithmic scale, indicating shot-noise-limited detection. The total number of electrons in a single pixel is nearly 10 ke-. (c) Photothermal SNR calculated for different number of averaged frames (\textit{N}). The fitted curve ($\alpha N^{0.505}$) has a slope of 0.505 in logarithmic scale, indicating mechanically very stable signal detection. Photothermal image acquisition time: 10 s (1000 frames). IR power: 6.05 mW @ 1729~cm\textsuperscript{-1}. Scale bar: 20~µm}
	\label{fig:fig2}
	\end{center}
\end{figure}

\subsection{System validation}
To evaluate our system's performance, we first demonstrated the mid-IR absorption by a thin film of PMMA spin-coated on a silicon substrate. 
As shown in figure~\ref{fig:fig2}(a), the IR focus on the substrate was elongated along the horizontal axis due to the oblique IR illumination.
This provided nearly 30~µm $\times$ 60~µm of FOV for the photothermal detection. 
The IR focus can be adjusted to a more circular spot by defocusing the parabolic mirror's position, but the defocused IR focus could reduce the IR intensity and photothermal signal. 
For the sake of proof-of-concept experiments, we kept the IR focus smaller to achieve a high-SNR signal.
Note that, we cropped the rectangular image FOVs into squares to obtain symmetric data visualization throughout the manuscript.

Our photothermal microscopy in practice can achieve shot-noise-limited sensitivity. 
The noise-floor in the interferometric signal is dominated by the photon shot noise~\cite{iSCATReviewNL}.
However, the wide-field laser illumination creates heterogeneous background across the FOV. 
The interferometric contrast from nanoscale samples can be buried under such heterogeneity. 
Luckily, photothermal detection is immune to those variations, providing nearly background-free images~\cite{OrritPT}. 
Owing to the virtual lock-in camera detection described above, the interferometric photothermal image is obtained by subtracting images at hot and cold states. 
The dominant noise component in difference images is the shot-noise contribution from the background signal that is the reflected light. 
Although interferometric image background has contrast variation across the FOV in the order of a few percent, the shot-noise level variations become quite negligible owing to relatively small differences. 

We experimentally evaluated the noise floor in our imaging system. 
We measured the noise in differential frames of the same FOV captured at a various number of collected photons (\textit{P}). 
This minimized the background fluctuations due to the nonuniform laser illumination. 
Figure~\ref{fig:fig2}(b) shows our noise calculations in the logarithmic scale. 
The fitted curve as a function of \textit{P} was $\alpha P^{0.498}$ with a constant scaling factor of $\alpha$. 
The exponent of the curve was very close to the theoretical value of 0.5. 
This indicated that our camera grants the shot-noise-limited detection for sufficiently saturated frames. 
The modulation fraction described in equation~\ref{eq:PTmod} was in the orders of the camera's shot-noise limited contrast which is defined by $\mathrm{1/\sqrt{P}}$. 
Averaging of shot-noise-limited image frames can reduce the noise floor by a factor of the square root of the number of frames \textit{N}. 
Therefore, multiple difference images were averaged in practice to obtain the photothermal signal with good data fidelity. 
The PMMA film was a great experimental example to show SNR improvement via frame averaging since a single photothermal frame had an SNR of 1.2. 
The SNR was calculated by dividing the maximum signal by the background standard deviation outside the IR spot. 
As shown in figure~\ref{fig:fig2}(c), the experimental image SNR curve which scales with $\alpha N^{0.505}$ was consistent with the theoretical curve ($\alpha N^{0.5}$). 
This indicated that the photothermal SNR can be greatly improved by frame averaging with minimal mechanical noise during the acquisition. 
Furthermore, the SNR in shot-noise-limited measurements can be further improved by using CMOS cameras with large pixel-well-depth recently used in interferometry measurements~\cite{Rohit_PNAS_2020}. The use of these large pixel-well-depth cameras at high-frame rates would enable ultra-fast single nanoparticle fingerprinting. 

\begin{figure}[!ht]
	\begin{center}
	\includegraphics[width =.6 \textwidth]{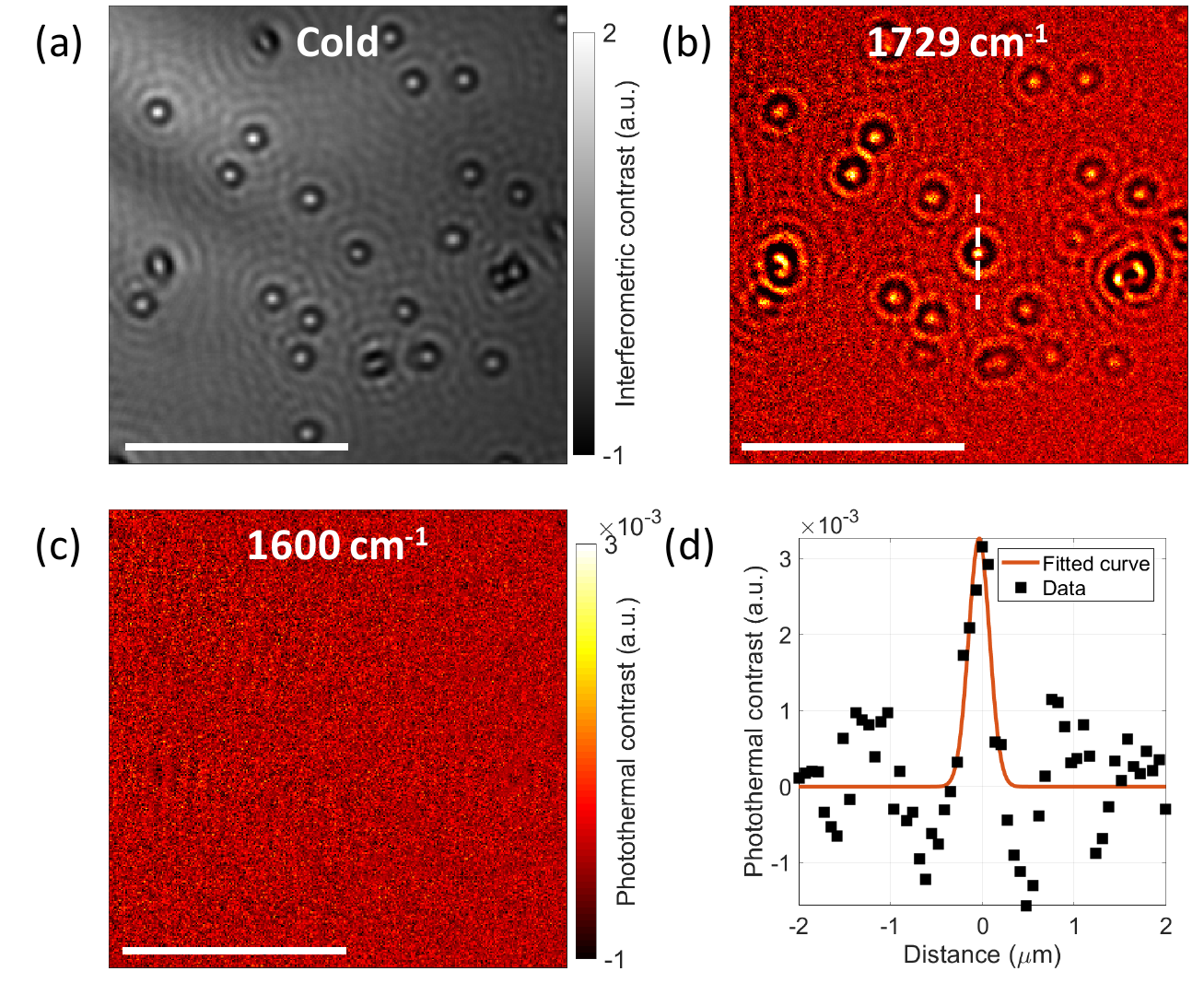}
	\caption{\textbf{Bond-selective interferometric imaging of 500~nm PMMA bead}. (a) Interferometric cold image. (b-c) Interferometric MIP images at C=O stretching absorption peak (1729~cm\textsuperscript{-1}) and off-resonance (1600~cm\textsuperscript{-1}), respectively. (d) Cross-section profile indicated by the white dashed line along the single bead in (b). The FWHM of the Gaussian fit is nearly 300~nm. The minimum and maximum points coincide at 500~nm separation ($\Delta_{feature} = $ 500~nm). Beads at the interface appear to be Gaussian-shaped objects with ($\Delta_{bead} = $ 350~nm), resulting in a deconvolved PSF with 357~nm FWHM ($\Delta_{PSF} = \sqrt{ \Delta_{feature}^2 - \Delta_{bead}^2} = $ 357~nm). Photothermal image acquisition time: 50 s. IR power: 2 mW @ 1729~cm\textsuperscript{-1} and 3.8 mW @ 1600~cm\textsuperscript{-1}. Scale bars: 10~µm}
	\label{fig:fig3}
	\end{center}
\end{figure}

Next, we perform proof-of-principle experiments on 500 nm PMMA nanospheres to demonstrate photothermal imaging of individual particles. 
The PMMA beads have a refractive index similar to biological nanoparticles, having similar particle polarizability for a given size (see equation~\ref{eq:polarizability}). 
Therefore, the interferometric contrast from PMMA beads is very close to those from the same size viruses or bacteria. 
Figure~\ref{fig:fig3}(a) shows the interferometric image from a cropped FOV, showing nearly a contrast level of 2. 
To obtain chemical imaging of these beads, we target the carboxyl group around the C=O stretching absorption peak at 1729~cm\textsuperscript{-1}. 
The resonance photothermal image has an SNR of 9. When the IR wavelength is tuned to off-resonance at 1600~cm\textsuperscript{-1}, we observe no photothermal contrast.
As we expected, the background heterogeneity due to laser illumination is greatly reduced in the photothermal images. 
This shows consistency with the differential images obtained under interferometric scattering microscopy~\cite{ISCATdifferential}. 
Furthermore, our photothermal measurements can achieve diffraction-limited lateral resolution in the visible spectrum (see figure~\ref{fig:fig3}(d)). 
According to Rayleigh's resolution definition, the first minimum of the cross-section profile across the 500~nm beads coincides with the peak value of around 500~nm. 
Under Gaussian object approximation~\cite{YilmazGaussianOBject}, the deconvolved PSF becomes 357~nm which is close to $\mathrm{\lambda /2NA}$.

\subsection{Biological sample characterization}
\begin{figure}[!ht]
	\begin{center}
	\includegraphics[width =.6 \textwidth]{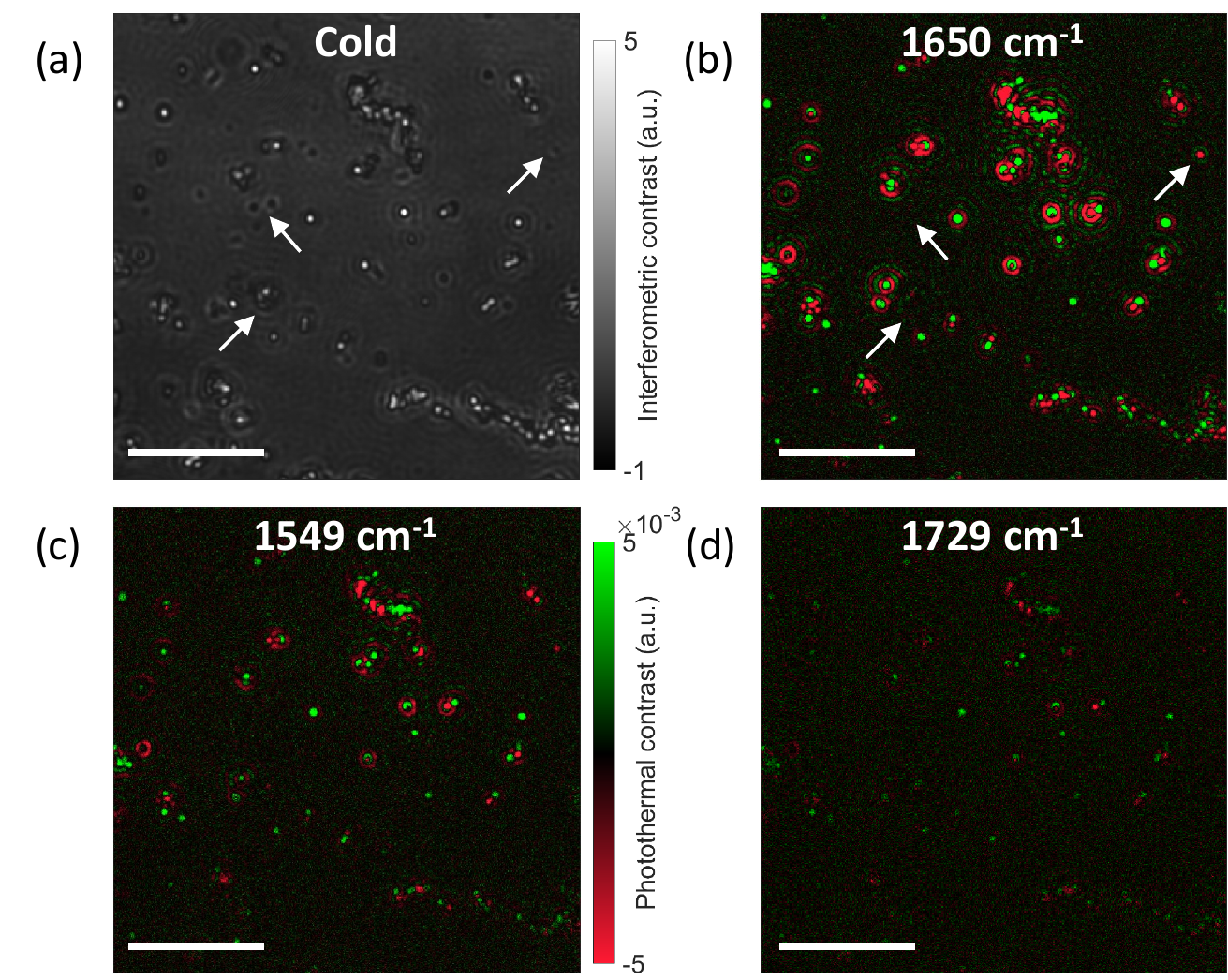}
	\caption{\textbf{Bond-selective interferometric imaging of \textit{S. aureus} bacteria.} (a) Interferometric cold image. (b-d) Multi-spectral MIP images at discrete wavenumbers; (b-c) Amide I (1650~cm\textsuperscript{-1}) and II (1549~cm\textsuperscript{-1}) bands, and (d) off-resonance (1729~cm\textsuperscript{-1}). The left two arrows in DC image (a) indicates sample like features which disappear in MIP images. The upper right arrow indicates that the negative interferometric contrast from \textit{S. aureus} due to its size appears to be negative in the MIP images. IR powers: 12.4 mW @ 1650~cm\textsuperscript{-1}, 9 mW @ 1549~cm\textsuperscript{-1}, and 6.05 mW @ 1729~cm\textsuperscript{-1}. Photothermal image acquisition time: 50 s (5000 frames). Scale bars: 10~µm}
    \label{fig:fig4}
	\end{center}
\end{figure}

To demonstrate bond-selective interferometric imaging of individual biological nanoparticles, we study bacteria and fungi as testing models. 
Figure~\ref{fig:fig4} shows multispectral imaging of spherical \textit{S. aureus} bacteria that are directly immobilized on the substrate (see Methods section). 
The interferometric image clearly shows their size variation across the FOV since the interferometric contrast is related to the particle's size.
To target major protein absorption peak in amide $\mathrm{\uppercase\expandafter{\romannumeral1}}$ band, the IR wavelength is tuned to 1650~cm\textsuperscript{-1}. 
The obtained photothermal image has a strong photothermal contrast with an SNR of about 60, indicating the rich protein content in \textit{S. aureus} cells. 
Furthermore, amide $\mathrm{\uppercase\expandafter{\romannumeral2}}$ band at 1549~cm\textsuperscript{-1} and off-resonance at 1729~cm\textsuperscript{-1} show the distinctive spectroscopic imaging at the same FOV. The weak-contrast at off-resonance wavelength comes from the residual IR absorption in the vibrational fingerprint region~\cite{IRaman}. 
Owing to the chemical specificity, photothermal detection is immune to the scattered signal from the unspecific immobilized particles.
The interferometric particle signal indicated by the left arrow in figure~\ref{fig:fig4}(a) disappears in the photothermal images, showing no contrast. 
The dark contrast particles (see e.g., the middle arrow), seem to be associated with unwanted signals based on the one-to-one comparison. 
However, the dark contrast particle indicated by the right arrow shows photothermal contrast. 
This result further emphasizes the chemical specificity of our technique which was inaccessible in the conventional interferometric microscopy. 
We also point out that the photothermal contrast could have a positive or negative sign. 
This stems from the fact that thermo-optic and thermal expansion coefficients are typically in opposite signs due to the particle density reduction~\cite{Masaru1ContrastMechanism}, counteracting each other according to the superposition in equation~\ref{eq:PTmod}. 
The overall photothermal effect on the signal strongly depends on the particle's size~\cite{Zhongming2017}. 
Either one of the coefficients could be dominant or they can cancel each other. 
Thus, some materials of a certain size could be invisible in this technique. 

\begin{figure}[!ht]
	\begin{center}
	\includegraphics[width =.6\textwidth]{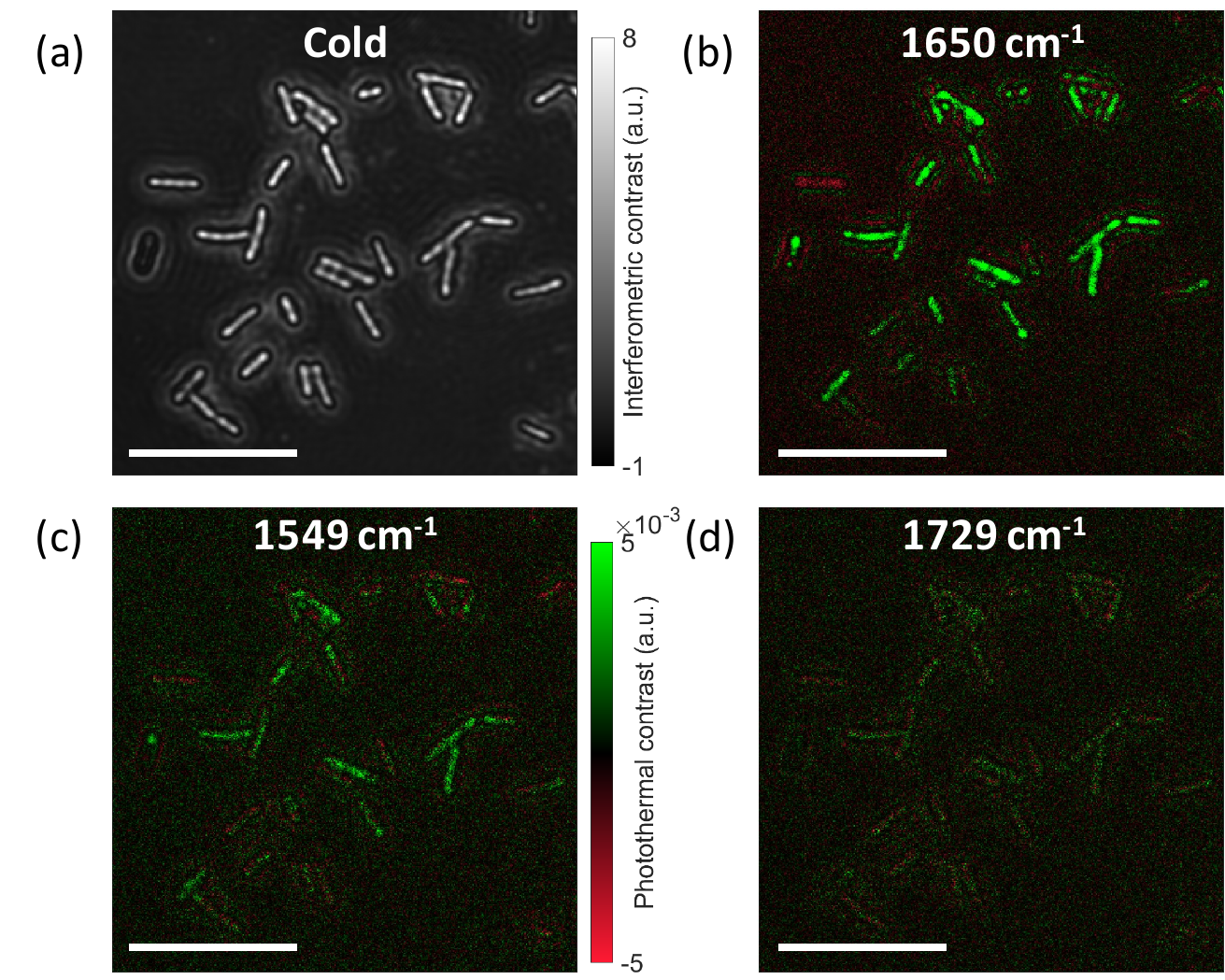}
	\caption{\textbf{Bond-selective interferometric imaging of \textit{E. coli} bacteria.} (a) Interferometric cold image. (b-d) Multi-spectral MIP images at discrete wavenumbers; (b-c) Amide I (1650~cm\textsuperscript{-1}) and II (1549~cm\textsuperscript{-1}) bands, and (d) off-resonance bond (1729~cm\textsuperscript{-1}). IR powers: 12.4 mW @ 1650~cm\textsuperscript{-1}, 9 mW @ 1549~cm\textsuperscript{-1}, and 6.05 mW @ 1729~cm\textsuperscript{-1}. Photothermal image acquisition time: 50 s (5000 frames). Scale bars: 10~µm}
    \label{fig:fig5}
	\end{center}
\end{figure}

Next, we image rod-shape \textit{E. coli} bacteria in the fingerprint region as shown in figure~\ref{fig:fig5}. 
Similar to the \textit{S. aureus} cells, the multispectral images are acquired at amide $\mathrm{\uppercase\expandafter{\romannumeral1}}$ (1650~cm\textsuperscript{-1}) and amide $\mathrm{\uppercase\expandafter{\romannumeral2}}$ (1549~cm\textsuperscript{-1}) bands and off-resonance bond (1729~cm\textsuperscript{-1}). 
The amide $\mathrm{\uppercase\expandafter{\romannumeral1}}$ band shows the strongest photothermal signal with an SNR of about 22, which is three-fold less than the \textit{S. aureus} cells. 
Nevertheless, \textit{E. coli} photothermal images also exhibit distinctive signal levels at the amide $\textrm{\uppercase\expandafter{\romannumeral2}}$ band and off-resonance bond which has much weaker photothermal contrast.
We also observe that one of the \textit{E. coli} cells have negative photothermal contrast although its interferometric image contrast is positive.
Interestingly, another \textit{E. coli} with negative interferometric contrast has a positive photothermal contrast. 
These results indicate that photothermal signal signs of the same particle types could be different even for similar interferometric contrast levels due to the differences in their size and shape. 
Together, these two distinctive bacteria results show the promise of this bond-selective interferometric scattering microscopy for high-throughput and multiplex differentiation of diverse bacteria populations.

\begin{figure}[!ht]
	\begin{center}
	\includegraphics[width =.8 \textwidth]{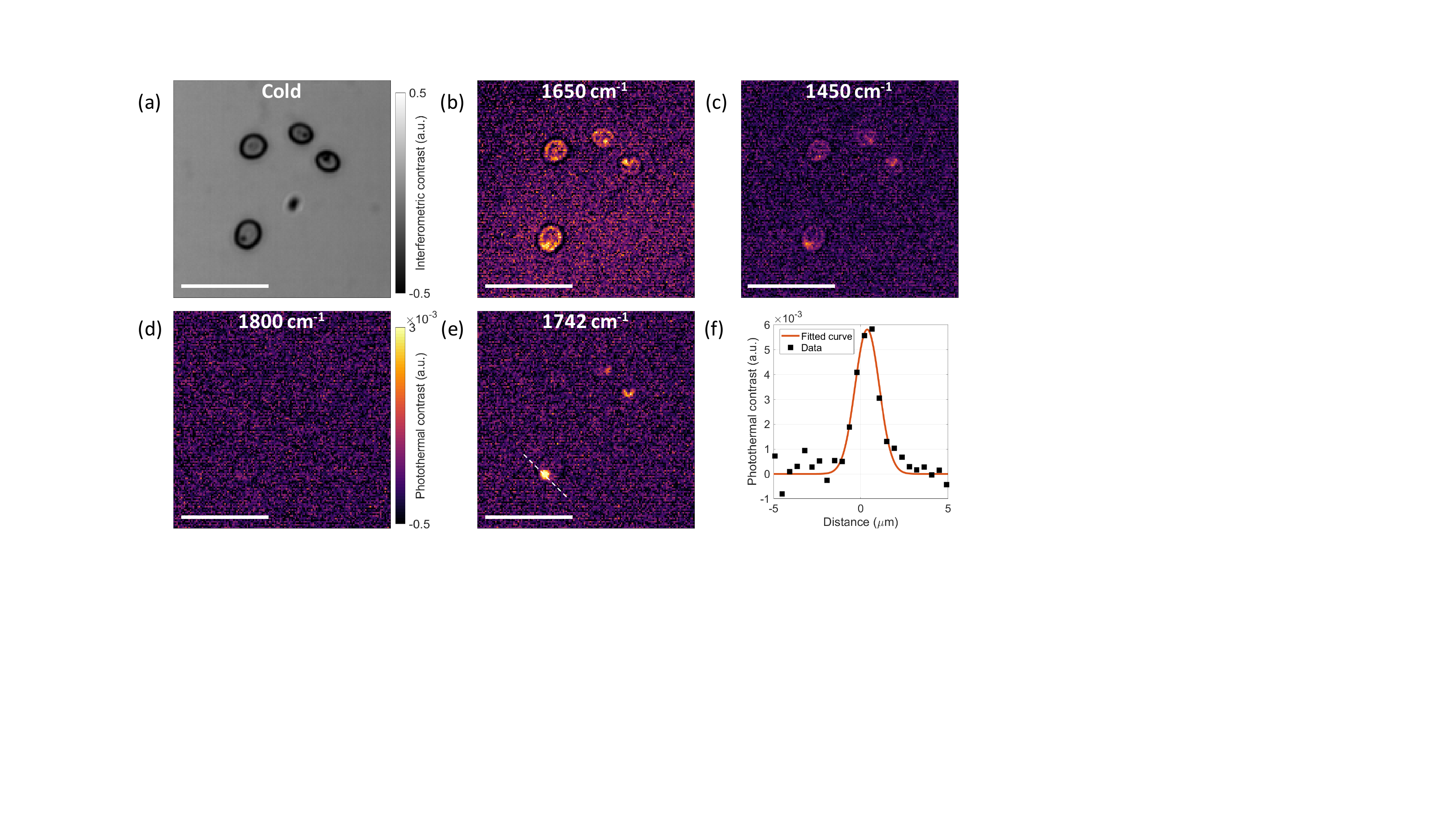}
		\caption{\textbf{Bond-selective interferometric imaging of live \textit{C. albicans} in phosphate buffer saline} (a) Interferometric cold image of \textit{C. albicans}. (b-d) Multi-spectral MIP images at discrete wavenumbers; (b) protein Amide I band (1650~cm\textsuperscript{-1}), (c) CH\textsubscript{2} bending (1450~cm\textsuperscript{-1}) in lipid and protein bonds, (d) off-resonance (1800~cm\textsuperscript{-1}), and (e) C=O stretching (1742~cm\textsuperscript{-1}) in phospholipid esters. (f) Cross-section profile indicated by the white dashed line along the lipid droplet in (e). The FWHM of the Gaussian fit is 1.64~µm. IR powers: 11.35 mW @ 1650~cm\textsuperscript{-1}, 11 mW @ 1450~cm\textsuperscript{-1}, 8.7 mW @ 1800~cm\textsuperscript{-1}, 12.9 mW @ 1742~cm\textsuperscript{-1}. Photothermal image acquisition time: 1.6 s (1000 frames). Scale bar: 20~µm}
    \label{fig:fig6}
	\end{center}
\end{figure}

We further investigate the capability of our technique on micron-scale single cells. 
We image \textit{C. albicans} fungi that have oval morphology. 
Since these cells are several microns in diameter larger than diffraction-limited resolution, the temporal coherence of the laser source hinders visualizing inside the cells due to the speckle noise. 
To address this limitation, we replace the green laser illumination with a partially coherent blue light-emitting diode (LED). This greatly reduces the speckle noise as shown in figure~\ref{fig:fig6}. 
The LED illumination is achieved in K\"{o}hler configuration which provides source-free and highly uniform sample illumination. We note that the frame rate in this experiments is set to 1250 frame/s using a fast CMOS camera (IL5, Fastec Imaging). To measure the protein distribution inside the cells, we tune the IR wavelength to 1650~cm\textsuperscript{-1} at amide $\mathrm{\uppercase\expandafter{\romannumeral1}}$ band and 1450~cm\textsuperscript{-1} at the amide $\mathrm{\uppercase\expandafter{\romannumeral3}}$ band. 
The strong signal from protein bonds is obtained whereas the off-resonance image shows no contrast. To avoid the lipid absorption peak, the off-resonance wavelength is set to 1800~cm\textsuperscript{-1} in contrast to bacteria samples.
Furthermore, we target the phospholipid ester band at 1742~cm\textsuperscript{-1} to visualize the lipid droplet inside the fungi cells. 
The photothermal image reveals a micron-scale lipid droplet in one of the cells. 
The Gaussian fit to the lipid droplet indicated by the dashed line has an FWMH of 1.64~µm.
Our findings show great promise for the understanding of such sub-cellular organelles with high chemical specificity.

\section{Conclusion}

In summary, we report the bond-selective interferometric scattering microscopy that enables fingerprinting individual nanoparticles captured on the layered substrate. 
We utilize 70~nm Si\textsubscript{3}N\textsubscript{4}/Si substrate to increase interferometric contrast by reducing the reference background. 
The analytical framework of the interferometric photothermal signal mechanism is formulated. 
A direct relation between the interferometric contrast and the photothermal signal has been discussed. 
We demonstrate proof-of-principle biological experiments on \textit{S. aureus} and \textit{E. coli} bacteria and \textit{C. albicans} fungi. 
Our technique provides high-throughput molecular information beyond affinity-specific molecular specificity without any labeling in the vibrational fingerprint region. 
Moreover, this direct chemical imaging method could open up exciting possibilities for the functional analysis of biological nanoparticles at a single-particle level.

Our study is limited to wavelength scale nanoparticles since photothermal imaging of sub-100~nm particles requires a much shorter pump and probe pulses as a consequence of the direct relationship between heat dissipation and particle size.
As substantiated by the experimental results and theoretical framework, we demonstrate that bond-selective interferometric microscopy can achieve chemical sensitivity for a broad range of nanoparticles currently studied by interferometric techniques. 
Although our mid-IR photothermal technique has been demonstrated on the IRIS configuration with nitride-coated silicon substrates, it can be easily expanded to other interferometric microscopy approaches using transparent substrates including iSCAT and COBRI. 
In these configurations, standard glass substrates must be replaced by mid-IR transparent microscope slides such as CaF\textsubscript{2}. 
It should be noted that our bond-selective interferometric scattering microscopy method can be readily applied to the recent enhancements in the common-path interferometric microscopy techniques such as high-resolution with computational microscopy~\cite{yurdakulAN}, enhanced visibility with pupil engineering~\cite{oguzhanOptica, kukuraPupil, COBRIpupil}, nanoscale 3D localization with defocus signal~\cite{ISCAT3Dtracking}, and high precision molecular mass quantification~\cite{ISCATMass}. 
Such improvements advance the mid-IR photothermal microscopy field towards the chemical nanoscopy of individual biological nanoparticles including viruses, exosomes, and proteins. Overall, our refined bond-selective interferometric microscopy has a great promise for high-resolution and high-throughput spectroscopic imaging of a broad size range of biological nanoparticles without any labeling.

\section*{Acknowledgments}
We thank Dr. Meng Zhang for providing bacteria samples. This work was supported by the National Institutes of Health R35GM126223, R44EB027018, and R42CA224844 to J.X.C. C.Y and M.S.U acknowledge European Union’s Horizon 2020 Future and Emerging Technologies (No. 766466).

\clearpage

\bibliography{references}   
\bibliographystyle{unsrt}   
\end{document}